\documentclass{article}
\usepackage{authblk}
\usepackage{graphicx}
\usepackage{siunitx}

\title{Single-step fabrication of surface waveguides in fused silica with few-cycle laser pulses}

\author[1]{Federico J. Furch}
\author[1]{W. Dieter Engel}
\author[1]{Tobias Witting}
\author[1]{Armando Perez-Leija}
\author[1]{Marc J. J. Vrakking}
\author[1*]{Alexandre Mermillod-Blondin}

\affil[1]{Max-Born-Institut f\"ur Nichtlineare Optik und Kurzzeitspektroskopie, Max-Born-Stra\ss e, D-12489 Berlin, Germany}

\affil[*]{Corresponding author: mermillod@mbi-berlin.de}

\begin{document}

\maketitle

\begin{abstract}

Direct laser writing of surface waveguides with ultrashort pulses is a crucial achievement towards 
all-laser manufacturing of photonic integrated circuits sensitive to their environment.
In this Letter, few-cycle laser pulses (with a sub-10 fs duration) are used to produce subsurface waveguides 
in a non-doped, non-coated fused silica substrate.
The fabrication technique relies on laser-induced microdensification 
below the threshold for nanopore formation. 
The optical losses of the fabricated waveguides are governed 
by the optical properties of the superstrate. We have measured losses 
ranging from less than 0.1~dB/mm (air superstrate) up to 2.8~dB/mm when immersion oil is applied on top of the waveguide.

\end{abstract}


Among the several fabrication techniques that may be used to produce waveguides in glass substrates \cite{Righini2014}, fs-laser direct writing \cite{Davis1996} is especially attractive because it does not require a cleanroom environment, is cost-effective, and offers a high throughput \cite{Righini2014}. Fs-laser writing enables full 3d waveguide fabrication in crystals \cite{Chen2014}, polymers \cite{Baum2007} as well as glasses \cite{Florea2003}. Although research efforts have been mostly concentrated on volume microprocessing, the direct fabrication of surface waveguides would greatly extend the domain of applications of laser-written photonic integrated chips. 
The propagation of an optical field in a surface waveguide implies the presence of an evanescent wave which can be exploited for refractive index sensing \cite{Lapointe2014, Lapointe2015}, plasmonic excitation sensing \cite{Sepulveda2006} or Fourier-transform spectrometry \cite{Coarer2007}.
However, using fs-laser direct writing on or near the surface of the host material is challenging.
Attempts in non-optimized glasses have resulted in surface swelling \cite{Bhardwaj2004}, cracking \cite{Lapointe2014} and ablation \cite{Torchia2007}. In pure fused silica, the direct fabrication of near-surface waveguides with a good refractive index contrast has even been considered impossible \cite{Berube2016}. The main reason is that fs-laser direct writing relies on pulse-to-pulse heat accumulation to induce controlled, localized heating of the substrate. After cooling, high-refractive index regions with light-guiding capabilities appear. In this scheme, the magnitude of the stress load produced upon thermal relaxation is a major limitation. Recently, several routes have been explored
to enable surface waveguide photoinscription. One strategy consists in using toughened glasses as a host substrate \cite{Lapointe2014, Lapointe2015}. Another approach is to enhance the photosensitivity of the glass by doping with silver ions \cite{AbouKhalil2017}. A third method relies on suppressing the air/dielectric interface by bonding a thin glass on top of the sample \cite{Berube2016}. Because the bond is ensured by weak van der Waals attractive forces (only manifesting when the sample and the cover glass are in close contact), such a method might only be applicable to planar substrates.
In this Letter, we describe the direct fabrication of surface waveguides in fused silica with the help of few-cycle laser pulses. Our method does not rely on pulse-to-pulse heat accumulation, but instead is based on the type 1 laser-matter interaction regime \cite{Mishchik2013b} triggered in a grazing incidence irradiation scheme. Modifications induced in the type 1 regime exhibit a smooth, uniform and positive refractive index change $\Delta n$, in contrast to type 2 modifications which are characterized by an important birefringence due to the presence of periodic nanogratings in the irradiated region. Furthermore, we demonstrate the possibility to control the propagation losses  by playing on the refractive index of the superstrate which opens a route for the laser-assisted fabrication of non-hermitian microoptical systems \cite{El-Ganainy2019}.

\begin{figure}[h!]
\centering
\includegraphics[width=0.75\linewidth]{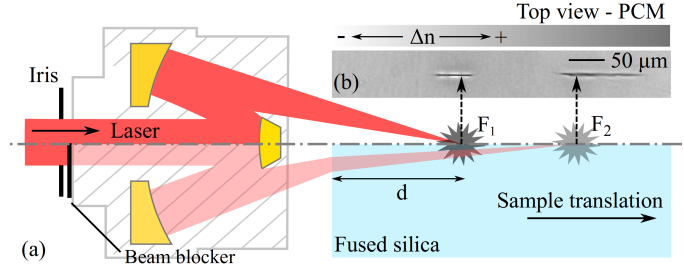}
\caption{(a): Cross-sectional view of the experimental setup used for laser microprocessing at grazing incidence. The part of the laser beam represented in bright red propagates through air only and forms the focus $F_1$. The beam blocker can be put in the beam path to prevent the formation of the aberrated focus $F_2$. (b): Phase-contrast microscopy (PCM) picture of the sample's top surface showing the refractive index distribution resulting from $F_1$ and $F_2$.} 
\label{fig:setup}
\end{figure}

The experimental setup is depicted in Fig.~\ref{fig:setup}. Few-cycle pulses (sub-10~fs duration, central wavelength at 800~\si{\nano\meter}) from a high repetition rate (400 kHz), non-collinear optical parametric amplifier \cite{Furch2016} are focused on the surface of a fused silica sample with a grazing incidence. The samples are parallelepipedic and polished to optical quality on all sides. In order to preserve the temporal structure of the laser pulse in the focal plane, we use a gold-coated reflective objective (numerical aperture 0.5) \cite{Piglosiewicz2011}.
Such an irradiation scheme provides two focii $F_1$ and $F_2$ formed by the upper and lower halves of the laser beam, respectively. 
The presence of a planar air/glass interface induces wavefront distortions and leads to the formation of an
aberrated focus $F_2$ \cite{Huot2007} which is shifted and spread along the propagation axis. The corresponding laser-induced modifications are shown in Fig.~\ref{fig:setup}(b). An analogous distortion happens in the time domain with consequences on the temporal profile of the irradiation \cite{Sun2015}. Because the spatio-temporal distortions imparted on the lower half of the beam increase with the amount of propagation in the sample, the irradiation characteristics vary when translating the sample along the propagation axis. In order to limit the irradiation to the part of the beam traveling through air only, the lower half of the laser beam was blocked. Furthermore, the entrance of the beam diameter was slightly reduced with the help of an iris, providing an overall transmission of 0.17 for the focusing unit. The fabrication of a single line-shaped microstructure is then straightforward. It suffices to irradiate the surface of the substrate continuously from edge to edge by translating the sample with the help of a stepper motor. In what follows, the laser pulse energy was constant with a value of $E = $ \SI{530}{\nano\joule} after the microscope objective and the speed of the stepper motor was \SI{60}{\micro\meter\per\second}. The laser-induced microstructures had a length of \SI{10}{\milli\meter} (i.e. the width of the sample). For higher pulse energies, intense laser light scattering occurred and the irradiated region exhibited a mix of negative and positive refractive index changes (not shown). These features are indicative of the type 2 interaction regime characterized by the formation of nanopores \cite{Bellouard2016, Mishchik2013b}.\\

\begin{figure}[h!]  
\centering
\includegraphics[width=0.5\linewidth]{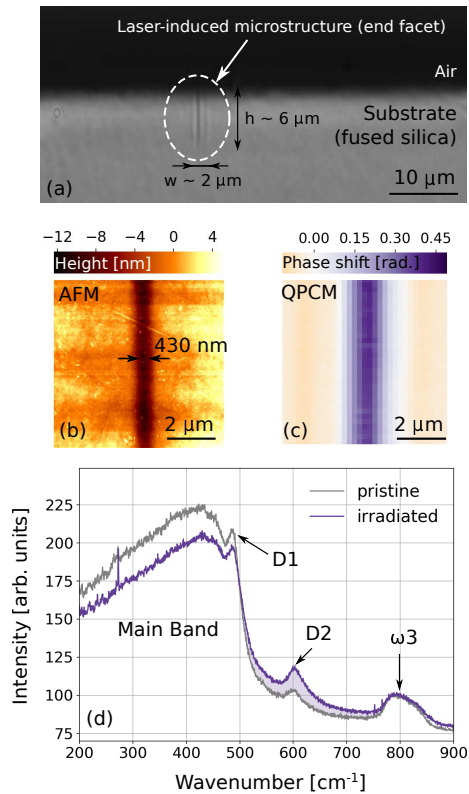}
\caption{Characterization of the laser-induced optical structures. (a): Side view of the sample acquired with an optical transmission microscope. The sample is illuminated with an halogen lamp. (b): Surface topography (top view) measured with an atomic force microscope (AFM).(c): Phase shift across the microstructure, measured with a spatial light interference microscope (SLIM). (d): Micro-Raman investigations of the irradiated zone (purple line) and of the pristine sample (grey line). The spectra were normalized with respect to the $\omega 3$ band.
} 
\label{fig:characterization}
\end{figure}

In Fig.~\ref{fig:characterization}(a) we show the laser footprint on the exit facet after sample polishing,
using optical transmission microscopy (numerical aperture 0.9). The laser-induced microstructure has a width $w \approx$~\SI{2}{\micro\meter} and a height $h \approx$~\SI{6}{\micro\meter}. Diagnostics of the top surface with an atomic force microscope (AFM) are presented in  Fig.~\ref{fig:characterization}(b), and reveal the presence of a shallow surface topography variation ($<$ \SI{-10}{\nano\meter}, about \SI{430}{\nano\meter} FWHM) on top of the laser-induced microstructure. The negative sign of the topology variation is indicative of a volume reduction (and hence a density increase) in the irradiated area, and is the opposite to what happens when microprocessing is performed in the type 2 regime, with longer (\SI{35}{\femto\second}) and more energetic (about \SI{2}{\micro\joule}) laser pulses where surface swelling as high as \SI{+250}{\nano\meter} was measured \cite{Bhardwaj2004}. An inversion in the sign of the surface topography is consistent with recent observations reporting a volume reduction of glass cantilevers irradiated in the type 1 regime and a net volume increase of cantilevers irradiated in the type 2 regime \cite{Bellouard2016}. We emphasize that the absence of material re-deposition in the AFM pictures hints towards a purely non-ablative process. 
The phase shift distribution $\Delta \phi$  across the laser-induced  microstructure presented in Fig.~\ref{fig:characterization}(c) was measured by spatial light interference microscopy \cite{Wang2011}. 
As expected from a local density increase, $\Delta \phi$ is positive in the irradiated volume \cite{Tan1998}. The corresponding spatial average of the laser-induced refractive index change $\overline{\Delta n} = \frac{\lambda_c \Delta\phi}{2\pi h} \approx 0.006$, where $\lambda_c =$\SI{550}{\nano\meter} is the central wavelength of the illumination source (an halogen light bulb in our case), $h =$~\SI{6}{\micro\meter} is defined in Fig.~\ref{fig:characterization}(a) and $\Delta\phi=$~\SI{0.43}{\radian} is 
the phase shift measured at the center of the microstructure. We emphasize that the magnitude of $\overline{\Delta n}$ obtained exceeds the value of $10^{-4} - 10^{-3}$ usually measured in the bulk for type 1 modifications \cite{Mishchik2013b}.\\

To further confirm our conclusion on the laser-induced compaction and refractive index increase, we have examined the signature of the fs laser-induced structural modifications in micro-Raman spectra \cite{Mishchik2013b, Shcheblanov2018} measured in the pristine and irradiated material, as shown in Fig.~\ref{fig:characterization}(d). The excitation wavelength was \SI{442}{\nano\meter} and the microscope objective used for excitation and collection of the Raman signal has a numerical aperture of 0.8 which provides a depth resolution of \SI{0.85}{\micro\meter}. The spectra were normalized with respect to the amplitude of the $\omega 3$ band \cite{Saavedra2014}. They reveal an increase of the D2 peak (centered at \SI{600}{\per\centi\meter}) in the irradiated volume, confirming a local laser-induced compaction \cite{Bellouard2008, Bellouard2016}. 
These micro-Raman measurements were carried out on the top surface of the sample and might not necessarily correspond to the maximum of compaction, presumably located in the center of the laser-induced microstructure (i.e. $\approx$~\SI{3}{\micro\meter} away from the surface).\\
\begin{figure}[h!]  
\centering
\includegraphics[width=0.6\linewidth]{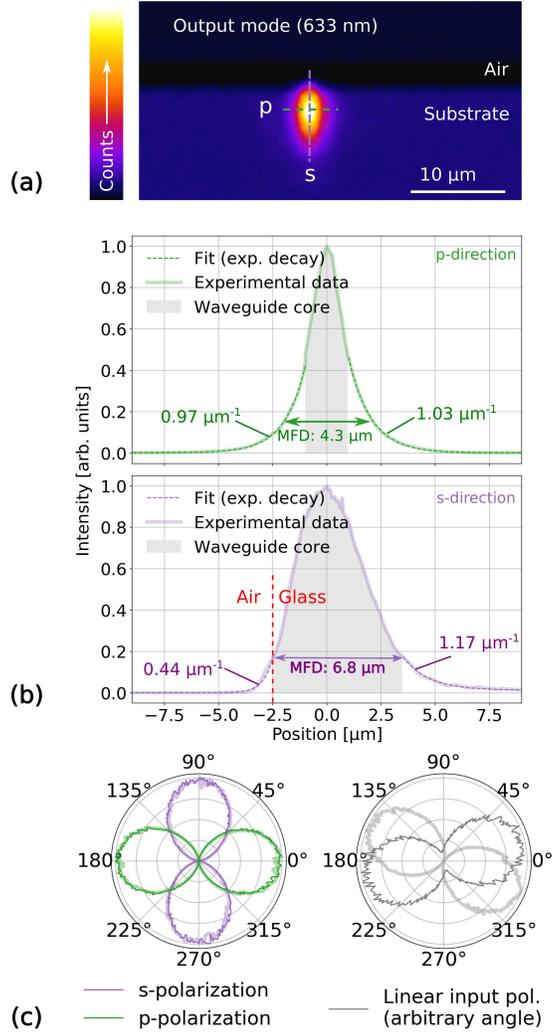}
\caption{(a): Near-field intensity profile at the exit of the laser-induced microstructure at a wavelength of \SI{633}{\nano\meter}. (b): Mode-field analysis. The regions in gray correspond to the size of the waveguide core deduced from Fig. 2(a). The dotted lines are first-order exponential fits of the experimental data. MFD: mode field diameter defined as the $1/e^2$ decay of the maximum intensity. (c): Transmitted intensity as a function of the analyzer angle for linear input polarizations in the p- and s- directions (left) and for an arbitrary linear input polarization (right). The dark thin lines represent the polarization of the beam after propagation in the waveguide and the thick, semi transparent lines represent the polarization of the input beam.
} 
\label{fig:mode}
\end{figure}
In order to examine the ability of the microstructures to guide light at optical frequencies, the fundamental mode of a CW He-Ne beam ($\lambda =$ \SI{633}{\nano\meter}) was focused onto the entrance facet of the laser-induced microstructures with a microscope objective (numerical aperture~0.42). A second microscope objective (Olympus MPlan, 100x, numerical aperture 0.9) used in combination with a tube lens (focal length of \SI{200}{\milli\meter}) formed a 111-fold magnified image of the end facet on a camera sensor.
Figure \ref{fig:mode}(a) shows the obtained output intensity distribution. It demonstrates that these laser-induced microstructures support optical waveguiding at \SI{633}{\nano\meter}.
A modal analysis in the directions parallel ($p-$) and perpendicular ($s-$) to the surface of the sample (see Fig.~\ref{fig:mode}(b)) provides a mode field diameter (MFD, defined as the $1/e^2$ decay of the maximum mode intensity), of 4.3 and 6.0 \si{\micro\meter} in the $p-$ and $s-$ directions, respectively. Out of the center region, the guided mode intensity decays exponentially \cite{Hu2009} with decay constants of $\approx$~\SI{1.1}{\micro\meter} and \SI{0.4}{\micro\meter} in glass and in air, respectively. We checked that the optical transfer function of the microscope objective did not significantly influence these values by applying a deconvolution algorithm to the curves shown in Fig.~\ref{fig:mode}(b). The point spread function used for the deconvolution was estimated numerically based on the model of Gibson and Lanni \cite{Gibson1989, Li2017}.

\begin{figure}[h!]
\centering
\includegraphics[width=0.6\linewidth]{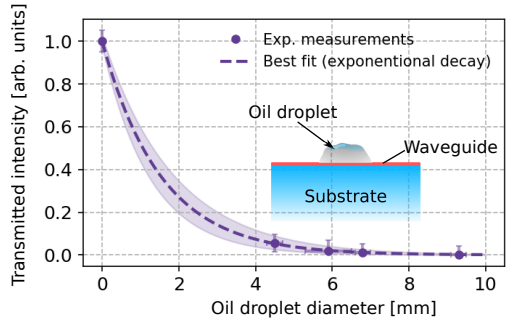}
\caption{Evolution of the optical transmission at \SI{633}{\nano\meter} as a droplet of immersion oil ($n_{oil}$ = 1.516) with a variable diameter is placed on top of the laser-induced surface waveguide.
The dotted line represents a fit of the experimental data using a single exponential decay. The presence of oil induces an attenuation of 2.83$^{+0.48} _{-0.66}$~\si{\decibel\per\milli\meter}.}
\label{fig:attenuation}
\end{figure}

The influence of the input polarization was examined by placing the waveguide between a polarizer and an analyzer. The transmission of the waveguide was measured as a function of the relative angle between the polarizer and the analyzer [see Fig.~\ref{fig:mode}(b)]. The polarization is maintained for input fields with a linear polarization in the s- and p- directions (see thick transparent lines in Fig.~\ref{fig:mode}(b) left). However, an input field with a linear input polarization in another direction becomes elliptically polarized upon propagation in the waveguide, which indicates that s- and p- polarized fields have different propagation constants.\\
Having confirmed the waveguiding capabilities of the laser-induced microstructures and their polarization-maintaining properties, we now examine the possibility to control the 
propagation losses by varying the refractive index of the superstrate $n_s$.
By applying the optimum end-fire coupling method \cite{Haruna1992}, losses $<$~\SI{0.1}{\decibel\per\milli\meter} were measured for $n_s = 1.00$~(in air).
The losses obtained when applying a drop of immersion oil on top of the waveguide are shown in Fig.~\ref{fig:attenuation}.
The diameter of the oil droplet was controlled by using a graduated microsyringe.\\
A first order exponential fit of the experimental data indicates that the superstrate-induced leakage is as strong as $2.83^{+0.48}_{-0.66}$~\si{\decibel\per\milli\meter}. 
The uncertainties were obtained from fits of the lower and upper bounds of the experimental values. In the so-called ray-optic approach of guided mode theory \cite{Ulrich1971}, the light propagating through a waveguide is described as a sum of oblique rays experiencing total internal reflection at the boundary of the waveguide. Changing $n_s$ changes the conditions for total internal reflection. When $n_s$ increases, the minimum angle for total internal reflection decreases and the steepest rays escape the waveguide \cite{Rehouma1994, Lapointe2015}.

The large attenuation coefficient deduced from the numerical fit indicates that the optical structures are sensitive to the refractive index of their environment and can thus be employed as refractive index sensors, with significant potential for lab-on-chip applications. These waveguides may for instance constitute the sensitive part of optical biosensors microsystems that are used for label-free bio-sensing \cite{Sepulveda2006}. The optical structures presented in this Letter may also be used as efficient photon/plasmon couplers to interface photonic and plasmonic architectures \cite{Guo2009}, or as the backbone for the fabrication of compact stationary-wave integrated Fourier-transform spectrometers \cite{Coarer2007}. Furthermore, the ability to control the losses in integrated waveguide configurations opens the door to new perspectives to manufacture non-Hermitian non-resonant photonic systems in connection with exceptional points singularities \cite{Miri2019}.

In this Letter we have presented a method to inscribe waveguides directly on the surface of a pure fused silica substrate. For the results presented here, few-cycle pulses from a high repetition rate non-collinear optical parametric amplifier have been utilized. The extent to which longer pulses can be utilized will be the subject of future investigations. We note that stretching the pulse by linear dispersion is not a viable option. The highly structured ultra broadband spectrum quickly leads to a multi-pulse structure in the time domain, which could potentially change the dynamics of the ionization process. We also note that new trends in non-linear pulse compression have the potential to bring few-cycle pulse capabilities to lasers typically used in laser material processing \cite{Lavenu2018}.

The results presented in this Letter represent the first demonstration of waveguides that are directly photoinduced on the surface of fused silica without the need for pre- (e.g. deposition of a photosentive material or applicaton of a cover slip) or post-processing of the target substrate.
The optical structures produced correspond to type 1 modifications, support waveguiding at optical frequencies, possesses polarization-maintaining properties and  exhibit a core refractive index change of $ \Delta n \approx + 0.006$ on average. AFM measurements and Micro-Raman investigations indicate that laser-induced microcompaction is at the origin of the observed refractive index change. 
The waveguides are sensitive to their environment, extending the capability of the direct laser write method to the rapid prototyping of compact optical non-Hermitian microsystems taking advantage of all the well-known benefits (small size and weight, low power consumption, improved reliability and vibration sensitivity \cite{Hunsperger2009}) of integrated optical devices.

\section*{Funding Information}
Deutsche Forschungsgemeinschaft, Grants ME4427/1-1 and ME4427/1-2.

\section*{Acknowledgments}
The authors thank J. Tomm and S. Schwirzke-Schaaf for their assistance with the micro-Raman measurements.

\section*{Supplemental Documents}






\bibliographystyle{plain}
\bibliography{swgs2019}



\end{document}